\documentclass[conference,a4paper]{IEEEtran}
\IEEEoverridecommandlockouts
\usepackage{cite}
\usepackage{amsmath,amssymb,amsfonts}
\usepackage{algorithmic}
\usepackage{graphicx}
\usepackage{textcomp}
\usepackage{xcolor}

\usepackage{multirow}
\usepackage{booktabs}
\usepackage{colortbl}
\def\BibTeX{{\rm B\kern-.05em{\sc i\kern-.025em b}\kern-.08em
    T\kern-.1667em\lower.7ex\hbox{E}\kern-.125emX}}
\begin{document}

\title{Encoder-Decoder-Based Intra-Frame Block Partitioning Decision\\
}

\author{\IEEEauthorblockN{Yucheng Jiang\IEEEauthorrefmark{1}\IEEEauthorrefmark{2}, Han Peng\IEEEauthorrefmark{2}, Yan Song\IEEEauthorrefmark{2}, Jie Yu\IEEEauthorrefmark{1}, Peng Zhang\IEEEauthorrefmark{2} and Songping Mai\IEEEauthorrefmark{1}}
\IEEEauthorblockA{\IEEEauthorrefmark{1}Shenzhen International Graduate School, Tsinghua University, Shenzhen, China}
\IEEEauthorblockA{\IEEEauthorrefmark{2}Peng Cheng Laboratory, Shenzhen, China}}

\maketitle

\begin{abstract}
The recursive intra-frame block partitioning decision process, a crucial component of the next-generation video coding standards, exerts significant influence over the encoding time. In this paper, we propose an encoder-decoder neural network (NN) to accelerate this process. Specifically, a CNN is utilized to compress the pixel data of the largest coding unit (LCU) into a fixed-length vector. Subsequently, a Transformer decoder is employed to transcribe the fixed-length vector into a variable-length vector, which represents the block partitioning outcomes of the encoding LCU. The vector transcription process adheres to the constraints imposed by the block partitioning algorithm. By fully parallelizing the NN prediction in the intra-mode decision, substantial time savings can be attained during the decision phase. The experimental results obtained from high-definition (HD) sequences coding demonstrate that this framework achieves a remarkable 87.84\% reduction in encoding time, with a relatively small loss (8.09\%) of coding performance compared to AVS3 HPM4.0.
\end{abstract}


\section{Introduction}

An efficient video coding standard is of great significance for the development of high-definition video industry. To this end, the next-generation video coding standards, Versatile Video Coding (VVC) and Audio Video coding Standard 3 (AVS3), were respectively released in July and November 2020. Compared to the widely used HEVC standard, both VVC and AVS3 achieves around 30\% performance improvement \cite{performance_comparison}, but at the cost of increasing the encoding complexity by an order of magnitude \cite{complexity_comparison}.

The intra-frame prediction in video coding is usually used for key frame compression. It utilizes spatial correlation between adjacent pixels to predict the current pixel value, thereby reducing the number of bits required to represent the current pixel value and thus reducing the amount of video data. Considering that the intra-frame block partitioning decision algorithm accounts for over 97\% of the intra-frame encoding time \cite{partition_complexity}, there has been a lot of research on reducing the time required for this decision making.

Traditional acceleration algorithms for intra-frame block partition usually use hand-crafted feature extraction for analysis, and skip some of the partitioning modes of the current coding unit (CU) based on analysis results \cite{tradition1,tradition2}. While compared with hand-crafted feature extraction, neural network-based acceleration algorithms can usually achieve higher coding efficiency. There are two primary categories of neural network acceleration algorithms. 

One category combines neural networks with traditional intra-frame prediction algorithms. These algorithms leverage the information acquired from the preceding neural network to reduce the number of traversals required by traditional intra-frame prediction algorithms. More specifically, a pooling-variable CNN was proposed to predict the continuation of current CUs partition in advance \cite{nn1}. Tech \emph{et al} proposed in their various works to use CNNs to predict different parameters related to the current LCU, which were utilized to constrain the block partitioning process \cite{nn2,nn3}. Xu \emph{et al} used two lightweight CNNs as classifiers, to distinguish whether the CU is to be partitioned and the direction in which the partition should take place \cite{nn4}. Moreover, one set of research employs CNNs to predict the partitioning probabilities of basic edges within a LCU. The predicted probability vector is then exploited in subsequent block partitioning decisions to expedite the judgement process \cite{b3,b1}.

The other category decouples neural networks from traditional intra-frame prediction algorithms, and allows the neural network to directly output the block partitioning result of the current LCU. The traditional intra-frame prediction algorithm only needs to make mode decisions without block partitioning decisions. For instance, Li \emph{et al.} proposed a MSECNN, as described in \cite{nn6}, for predicting partitioning results in a greedy way. However, the proposed network architecture is excessively complex, leading to a significant increase in computational load. When selecting the Top-1 result, the method proposed in \cite{b1} can replace the block partitioning decision process, but it demonstrates a significant decrease in performance.

In this paper, we propose a novel approach utilizing a compact network structure to accurately predicte block partitioning results. By fully parallelizing the process of predicting block partition, significant reductions in encoding time can be achieved. The proposed network has been trained and validated within the AVS3 coding framework, but can be easily adapted to the VVC framework with minor modifications.

\section{Task Formulation}

For video coding standards such as VVC and AVS3, once the partition of the parent node has been determined in the CU block partitioning decision process, the decision order for its child nodes is fixed. Therefore, the block partitioning result of the LCU can be represented as a structured variable-length integer vector, with each integer element in the vector ranging from 0 to 5, corresponding to the six partitioning modes of Non-Split (NS), QuadTree (QT), Horizontal Binary-Tree (BT-H), Vertical Binary-Tree (BT-V), Horizontal Ternary-Tree (TT-H) and Vertical Ternary-Tree (TT-V) in the VVC standard, or NS, QT, BT-H, BT-V, Horizontal Extend Quad-Tree (EQT-H) and Vertical Extend Quad-Tree (EQT-V) in the AVS3 standard. 

\begin{figure}[tp]
\centering
\includegraphics[width=0.43\textwidth]{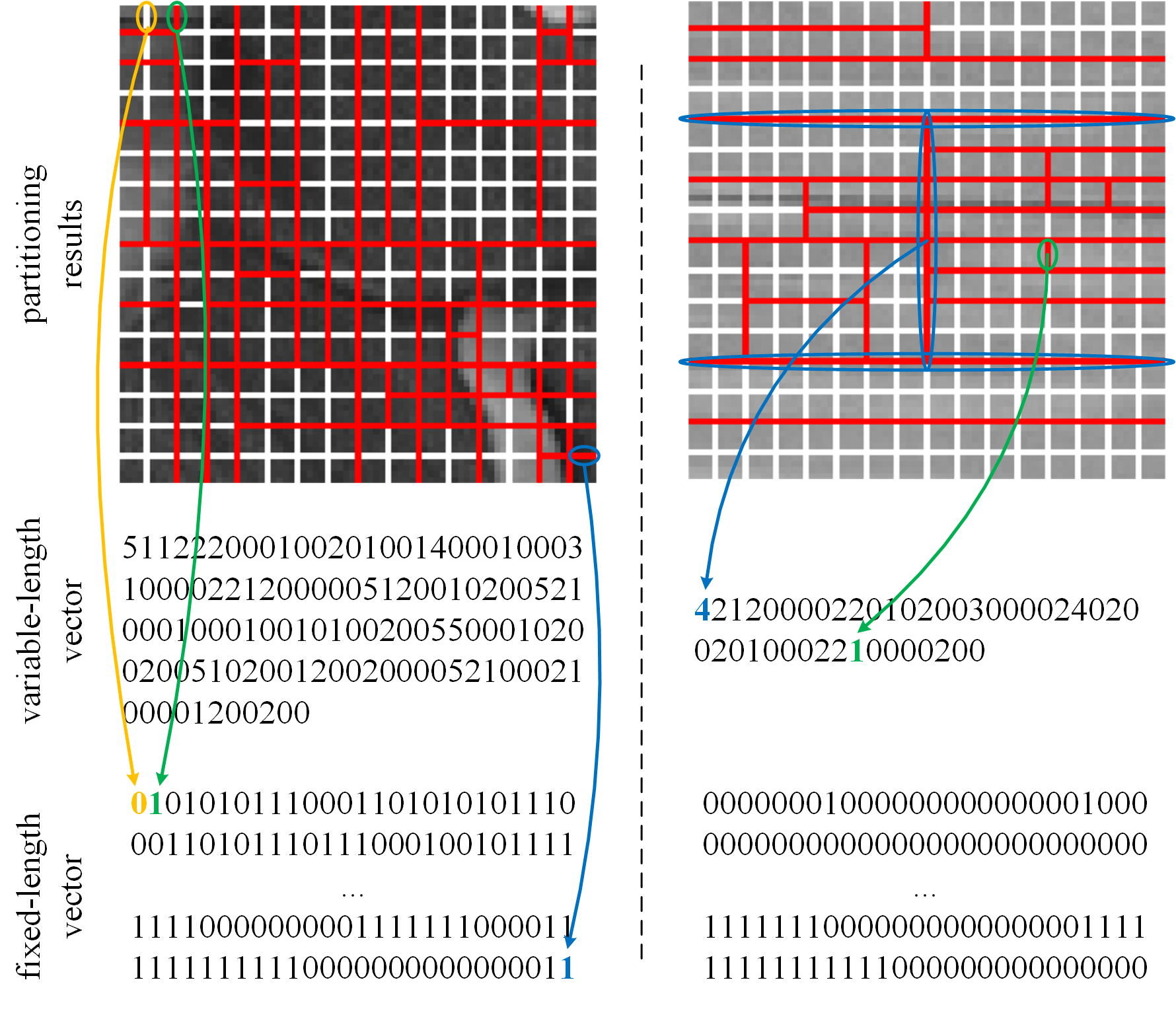}
\caption{Variable-length vector and Fixed-length vector representation of CTU partitioning results.}
\label{fig1}
\end{figure}

As described in \cite{b1}, the partitioning result of a 64x64-sized LCU can also be represented as a fixed-length vector consisting of 480 binary elements of 1 or 0, where 1 represents that the corresponding 4-pixel length edge (basic edge) belongs to the boundary of the partitioning block, and 0 represents the opposite. Fig.~\ref{fig1} provides a visual representation of these two different forms of representation discussed above, using two LCUs as examples. It should be noted that, in the actual inference process, the elements in the fixed-length vector are floating-point numbers ranging from 0 to 1, representing the probability of the basic edge belonging to the block partitioning boundary.

\begin{figure*}[tp]
  \centering
  \includegraphics[width=0.90\textwidth]{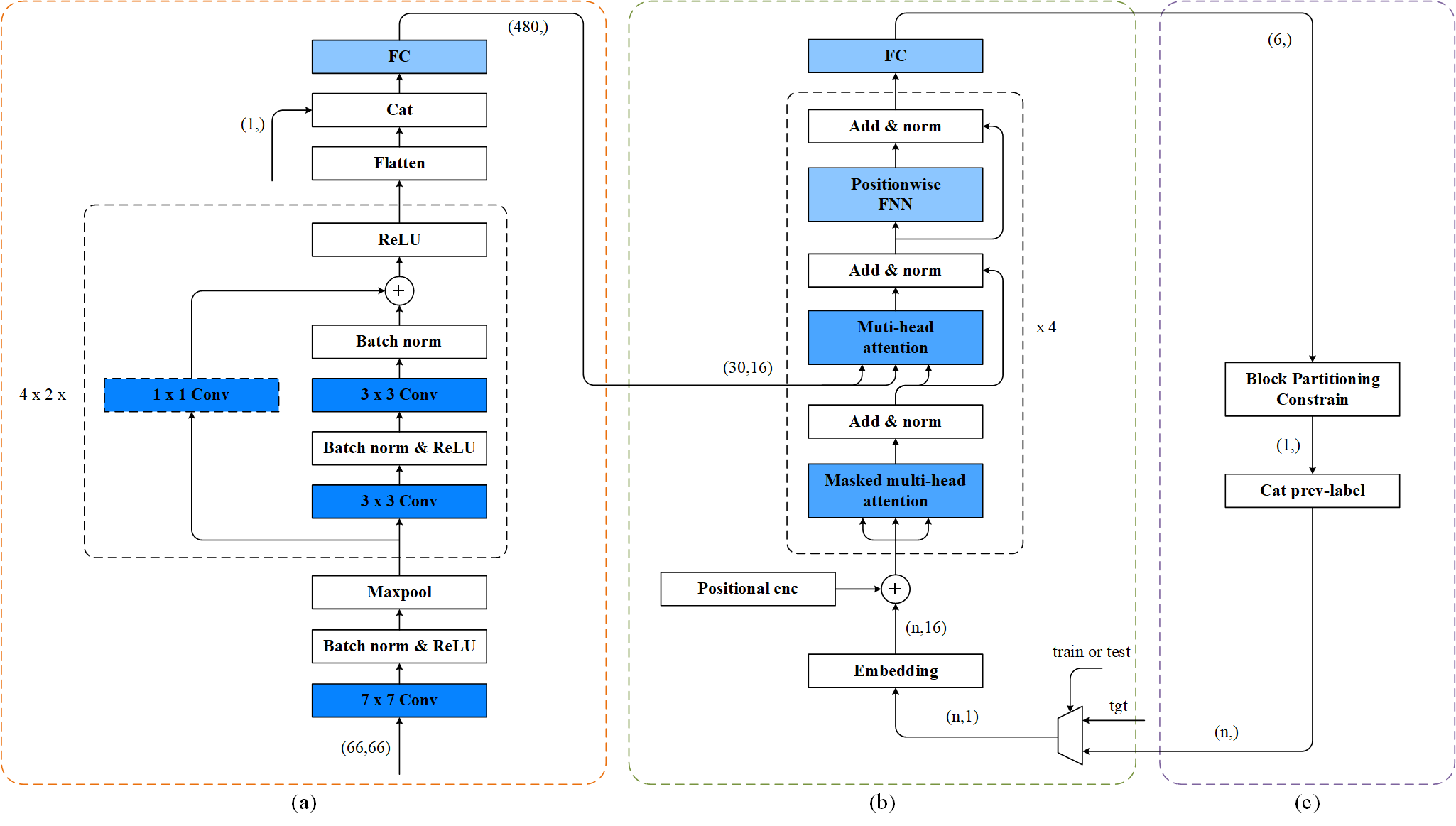}
  \caption{Neural network architecture. (a) CNN encoder. (b) Transformer decoder. (c) Block Partitioning Constrain module}
  \label{fig2}
\end{figure*}

Our objective is to use a neural network to predict the variable-length vector for each LCU. This vector can then be utilized by the video encoder to replace the recursive process of block partitioning decisions. The retention of only the mode decision algorithm leads to a substantial reduction in encoding time. Based on previous studies \cite{b1}, the pixels of a LCU can be encoded to a fixed-length vector using a neural network similar to ResNet. With this encoding, the subsequent prediction task can be reformulated as a sequence-to-sequence (seq2seq) task. In this task, the input sequence comprises flattened block partitioning information, while the output sequence consists of structured block partitioning information. In other words, we need to extract structured information from the fixed-length vector.

\section{Proposed Method}

Fig.~\ref{fig2} illustrates the neural network based on an encoder-decoder architecture proposed in this study. The CNN shown in subfigure (a) encodes the pixel data into a floating-point vector of length 480. Then the Transformer decoder in (b) performs the transformation from fixed-length vector to variable-length vector. The purpose of Block Partitioning Constrain (BPC) module in (c) is to ensure that the variable-length vector generated by the neural network conforms to the AVS3 coding constrains and can be correctly decoded.

Compared with the fixed threshold method in \cite{b3} and the decision tree (DT) method in \cite{b1}, the network structure proposed in this study has the following advantages,

\begin{itemize}

\item The self-attention mechanism in the Transformer decoder facilitates the integration of the previous block partitioning decision results, especially the partitioning information of the parent nodes. This correlation between different decision layers was usually overlooked in other methods. 

\item The decoder can automatically extract information by enhancing the attention of the current CU element in the encoder output feature. This replaces the manual construction of the DT model input. Furthermore, this structure achieves framework uniformity, eliminating the need to train different DT models for different CU sizes.

\item The variable-length vector output from the model can be directly used for LCU coding, completely replacing the recursive process of block partitioning decision making.

\end{itemize}

\subsection{CNN Encoder}

In the traditional AVS3 encoding algorithm, the optimal intra prediction mode for the current CU is computed by utilizing the reconstructed pixels from the top, left, and top-left neighboring CUs as references \cite{b2}. Therefore, we fold the reference pixels corresponding to each 64x64-sized LCU and fill them into the upper two rows and left two columns of a 66x66-sized block, while the remaining part is filled with the luminance pixels of the current LCU. We also input the Quantization Parameter (QP) into the neural network in addition to the pixel block, as QP value significantly affects block partitioning results.

The architecture of the CNN encoder is similar to the standard ResNet18 \cite{b4}, consisting of an input convolutional layer, four residual blocks, and an output fully connected layer. The difference lies in the size of the convolutional kernel in the convolutional layer and the dimension of the fully connected layer. Additionally, before entering the output fully connected layer, the QP value is concatenated with the flattened feature vector, enabling the probability vector of the model output to be affected by changes in the QP value.

\subsection{Transformer Decoder}

We employ the standard Transformer Decoder \cite{b5} for the seq2seq task, benefiting from its multi-head self-attention mechanism and multi-head encoder-decoder attention mechanism. These mechanisms enable us to effectively extract structured information from the fixed-length vector, enhancing the model's ability to capture and utilize contextual dependencies.

To prevent limited expressive capacity of the model caused by a small feature vector dimension, it is necessary to reshape the fixed-length vector. In consideration of the correlation typically found among basic edges located in the same row or column during block partition, and with each row or column comprising 16 basic edges, we reshape the fixed-length vector of length 480 into a (30,16) shape, thereby increasing the feature vector dimension from 1 to 16.

As shown in Fig.~\ref{fig2}(b), the decoder consists of 4 layers of Transformer decoder modules, with a hidden layer dimension of 16 and 4 attention heads. Its output is computed by (\ref{eq1}).

\begin{equation}
	\label{eq1}
  {{l}_{t}}=Decoder({{p}_{0}},{{p}_{1}},...,{{p}_{479}},{{l}_{0}},{{l}_{1}},...,{{l}_{t-1}})\in {{R}^{N}}
\end{equation}

\noindent where ${{p}_{i}} $ represents the i-th element of a fixed-length vector, and ${{l}_{j}} $ represents the output of the decoder at the j-th step. $N $ is the vocabulary size. For this task, since there are only 6 block partitioning modes in AVS3 and VVC, $N $ is set to 6. Specifically, the reshaped fixed-length vector is used as the key-value pairs and the concatenated historical output of the decoder is used as the query. The output vector of the decoder represents the probability of the 6 block partitioning methods corresponding to the current CU.

\subsection{Block Partitioning Constrain}

Multiple constraints exist for block partitioning decision of CUs in the AVS3 video coding standard. Failure to meet these constraints can result in a undecodable bitstream. Specifically, these constraints include: the aspect ratio of child CUs after partition cannot exceed 8, the total partitioning depth cannot exceed 6, non-QuadTree (QT) partitioning nodes cannot have child nodes that are QT partitioned, and so on. We design BPC module, which selects the block partitioning result that satisfies the encoder constraints from the probability vector output by the decoder. This is done by choosing the maximum probability value that meets the encoder constraint, rather than simply selecting the highest probability value. Since we adopt the teacher-forcing training method \cite{b5}, BPC module is not used during the training process. In this way, during both the training and testing processes, the model receives queries that adhere to encoding constraints. This ensures that the data is drawn from the same distribution.

\subsection{Neural Network Training}

We adjusted high-definition images from two publicly available datasets \cite{dataset1,dataset2} to HD resolution, and concatenated them into YUV video format. The resulting video consisted of a total of 19200 frames. Then we encoded the video using AVS3 reference software, and extracted the data. Each set of data consists of stitched 66x66 luminance pixel data, QP value, fixed-length vector labels, and variable-length vector labels. Additionally, over 85\% of the fixed-length vectors in the dataset have less than half of their elements equal to 1. To tackle this data imbalance and improve the accuracy of complex LCU block partition predictions, we balanced the dataset by adjusting the number of LCUs with deep and shallow partitions during training.

In contrast to \cite{cnn_transformer}, where a pre-trained CNN is required to provide input data for the Transformer decoder, our decoder can utilize fixed-length vector labels as input. This enables us to use a two-step training approach that includes both independent training and joint training. Independent training serves as a form of pre-training, supplying the initialization parameters for both the CNN encoder and the Transformer decoder. This methodology reduces the time required for joint training and subsequently enhances the overall predictive accuracy of the model.

\section{Experimental Results}

\subsection{Neural Network Prediction}

\begin{figure}[tp]
  \centering
  \includegraphics[width=0.355\textwidth]{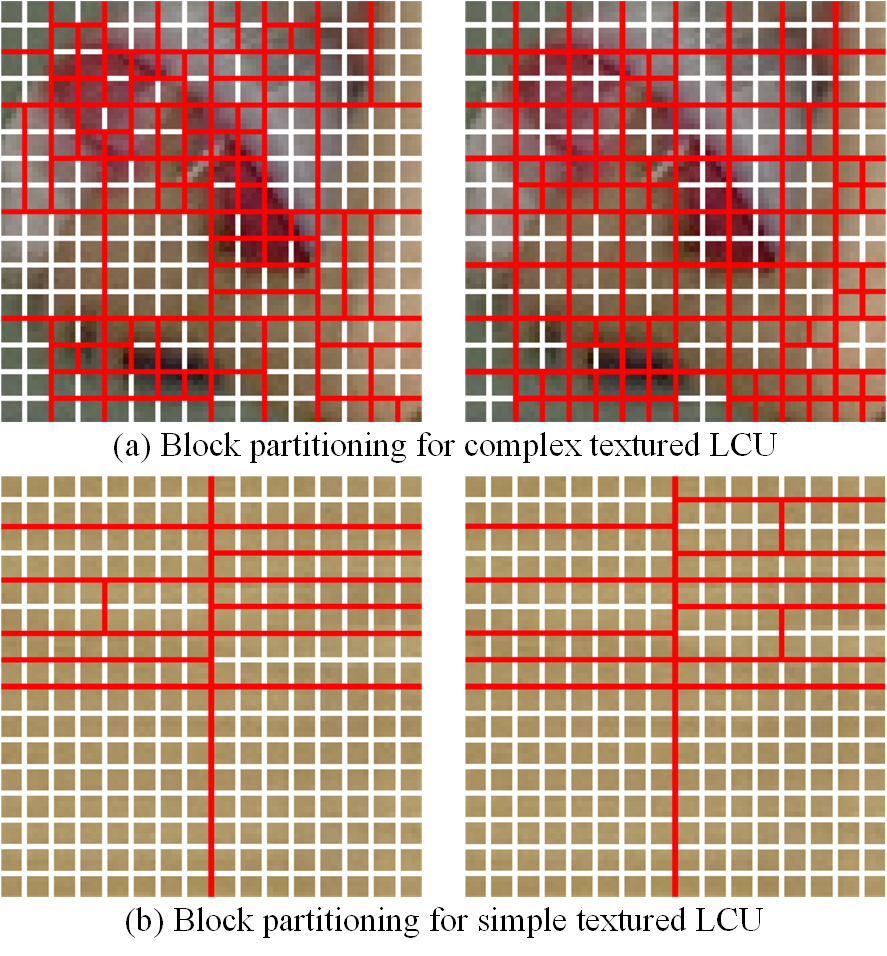}
  \caption{Comparison of block partitioning labels (left) and neural network prediction results (right) for LCU with different textured complexity levels}
  \label{fig3}
\end{figure}

Fig.~\ref{fig3} presents the prediction results of the trained network, which indicates that the network exhibits higher accuracy in predicting the partition of LCU blocks with simple textures, but lower accuracy in predicting more complex texture LCUs. Thus, improving the model's predictive capability on complex textures is crucial to further enhancing network performance. On average, the validation accuracy of the network can reach over 80\%, which is higher than the TOP1 accuracy of 67.24\% reported in \cite{b1} using the DT models.

\subsection{Video Coding Performance}

We deployed the trained neural network and AVS3 reference software on a heterogeneous platform to evaluate their coding performance and efficiency. The neural network was executed on a NVIDIA RTX3090 GPU, while the AVS3 reference software was executed on an Intel Xeon Gold 6148 CPU. It should be noted that to reduce training cost and time, our training, testing, and performance evaluation are all focused on HD video sequences. In the future, LCU data of videos with different resolutions will be added to the dataset to enhance the applicability of the network.

As shown in Table \ref{tab1}, we conducted joint testing of neural networks with reference software versions High-Performance Model (HPM) 14.0 and HPM4.0, and compared the test results with the coding results obtained solely using the reference software. The BDBR stands for Bjøntegaard delta bit rate \cite{bd_psnr}, while TS represents encoding time saving, which is calculated from (\ref{eq2}),

\begin{equation}
	\label{eq2}
  TS=\frac{{{T}_{HPM}}-\min ({{T}_{NN}},T_{HPM}^{'})}{{{T}_{HPM}}} 
\end{equation}

\noindent where ${{T}_{HPM}} $ represents the encoding time of the reference software, ${T_{HPM}^{'}} $ represents the encoding time of the reference software with block partitioning decisions removed, and ${{T}_{NN}} $ represents the inference time of the block partitioning results of the neural network. Due to the parallelized design, theoretically, the encoding time can be reduced by up to 97\%, which refers to the time required for block mode decisions. However, limited by our GPU performance, we save 93.97\% and 87.84\% of the encoding time for two versions of reference software respectively. 

In terms of coding performance, as shown in Table \ref{tab2}, the BDBR metric exhibits a decrease of 8.09\% compared to HPM4.0, which outperforms the 8.97\% performance decrease reported in \cite{performance_comp} using the same reference standard. Compared to the methods designed for VVC, our approach reduces more complexity, but its performance is inferior and requires further improvement.

\begin{table}[tp]
  \centering
  \caption{Performance of the proposed method in comparison with reference softwares in AI coding configuration}
    \begin{tabular}{cccccc}
    \toprule
    \multicolumn{2}{c}{\multirow{2}[4]{*}{Sequences}} & \multicolumn{2}{c}{HPM14.0} & \multicolumn{2}{c}{HPM4.0} \\
\cmidrule{3-6}    \multicolumn{2}{c}{} & BDBR  & TS    & BDBR  & TS \\
    \midrule
    \multicolumn{1}{c}{\multirow{5}[2]{*}{Class B}} & \multicolumn{1}{l}{BasketballDrive} & 13.73\% & 94.48\% & 3.44\% & 88.84\% \\
          & \multicolumn{1}{l}{BQTerrace} & 18.02\% & 93.40\% & 9.41\% & 87.38\% \\
          & \multicolumn{1}{l}{Cactus} & 16.22\% & 93.30\% & 10.01\% & 85.92\% \\
          & \multicolumn{1}{l}{MarketPlace} & 9.73\% & 95.45\% & 5.24\% & 91.18\% \\
          & \multicolumn{1}{l}{RitualDance} & 15.92\% & 93.21\% & 12.33\% & 85.88\% \\
    \midrule
    \multicolumn{2}{c}{Average} & 14.73\% & 93.97\% & 8.09\% & 87.84\% \\
    \bottomrule
    \end{tabular}%
  \label{tab1}%
\end{table}%

\begin{table}[tp]
  \centering
  \caption{Performance of the proposed method in comparison with state of the art methods in AI coding configuration}
    \begin{tabular}{ccccc}
    \toprule
    Standards & Versions & Methods & BDBR  & TS \\
    \midrule
    \multirow{2}[2]{*}{AVS3} & HPM4.0 & Ours  & 8.09\% & 87.84\% \\
          & HPM4.0 & Han, \textit{et al.} \cite{performance_comp}   & 8.97\% & 97\% \\
    \midrule
    \multirow{2}[2]{*}{VVC} & VTM7.0 & Li, \textit{et al.} \cite{nn6}   & 2.78\% & 65.75\% \\
          & VTM10.2 & Tissier, \textit{et al.} \cite{b1}   & 2.31\% & 74.60\% \\
    \bottomrule
    \end{tabular}%
  \label{tab2}%
\end{table}%

\section{Conclusion}

This paper proposes an encoder-decoder structured neural network for predicting the intra-frame block partitioning results of the LCUs. Firstly, the CNN encoder extracts local features from the LCU pixels. Then, a Transformer decoder is incorporated to predict the block partitioning results. Compared to other algorithms such as DT, the proposed structure achieves superior prediction accuracy. Furthermore, our designed BPC module enables the output of the Transformer decoder to be directly utilized by the video coding frameworks. Therefore, by adopting a parallelized design, this heterogeneous framework can save a significant amount of encoding time. Due to the similarity of the intra-frame block partitioning decision algorithm, the model can be easily adapted to other video coding algorithms, such as VVC, through retraining while maintaining the same network structure.

\end{document}